\documentclass[twocolumn,letterpaper,aps,prl,superscriptaddress,showpacs,nofootinbib,floatfix]{revtex4}

\usepackage{graphicx}
\usepackage{slashbox}
\usepackage{color}
\usepackage{brackets}
\usepackage{psfrag}
\usepackage{tabularx}
\usepackage{comment}
\begin{document}

\ProvideTextCommandDefault{\textonehalf}{${}^1\!/\!{}_2\ $}

\title{First Measurement of the Asymmetry and the Gerasimov-Drell-Hearn Integrand from $\vec{{^3}He}(\vec{\gamma},p)d$ reaction at  the Incident Photon Energy of 29 MeV}

\author{G.~Laskaris}\email[Electronic address: ]{gl41@duke.edu}
\affiliation{Triangle Universities Nuclear Laboratory, Durham, North Carolina 27708, USA}
\affiliation{Department of Physics, Duke University, Durham, North Carolina 27708, USA}
\author{W.~Ji}
\affiliation{Tsinghua University, Beijing, 100084, China}
\author{X.~Yan}
\affiliation{Triangle Universities Nuclear Laboratory, Durham, North Carolina 27708, USA}
\affiliation{Department of Physics, Duke University, Durham, North Carolina 27708, USA}
\author{J.~Zhou}
\affiliation{Triangle Universities Nuclear Laboratory, Durham, North Carolina 27708, USA}
\affiliation{Department of Physics, Duke University, Durham, North Carolina 27708, USA}
\author{W.~R.~Zimmerman}
\affiliation{Triangle Universities Nuclear Laboratory, Durham, North Carolina 27708, USA}
\affiliation{Department of Physics, Duke University, Durham, North Carolina 27708, USA}
\author{M.~W.~Ahmed}
\affiliation{Triangle Universities Nuclear Laboratory, Durham, North Carolina 27708, USA}
\affiliation{Department of Physics, Duke University, Durham, North Carolina 27708, USA}
\affiliation{Department of Mathematics and Physics, North Carolina Central University, Durham, North Carolina 27707, USA}
\author{T.~Averett}
\affiliation{College of William and Mary, Williamsburg, Virginia 23187, USA}
\author{A.~Deltuva}
\affiliation{Centro de F\'{i}sica Nuclear da Universidade de Lisboa, P-1649-003 Lisboa, Portugal}
\author{A.~C.~Fonseca}
\affiliation{Centro de F\'{i}sica Nuclear da Universidade de Lisboa, P-1649-003 Lisboa, Portugal}
\author{H.~Gao}\email[Electronic address: ]{haiyan.gao@duke.edu}
\affiliation{Triangle Universities Nuclear Laboratory, Durham, North Carolina 27708, USA}
\affiliation{Department of Physics, Duke University, Durham, North Carolina 27708, USA}
\author{J.~Golak}
\affiliation{M. Smoluchowski Institute of Physics, Jagiellonian University, PL-30059 Krak\'{o}w, Poland}
\author{A.~Kafkarkou}
\altaffiliation{deceased, 16$^{th}$ of February 2014.}
\affiliation{Triangle Universities Nuclear Laboratory, Durham, North Carolina 27708, USA}
\affiliation{Department of Physics, Duke University, Durham, North Carolina 27708, USA}
\author{ H.~J.~Karwowski}
\affiliation{Triangle Universities Nuclear Laboratory, Durham, North Carolina 27708, USA}
\affiliation{Department of Physics and Astronomy, University of North Carolina at Chapel Hill, Chapel Hill, North Carolina 27599, USA}
\author{B.~Lalremruata}
\affiliation{Mizoram University, Aizawl 796004, India}
\author{J.~Manfredi}
\affiliation{NSCL, Michigan State University, East Lansing, Michigan 48824, USA}
\author{J.~M.~Mueller}
\affiliation{Triangle Universities Nuclear Laboratory, Durham, North Carolina 27708, USA}
\affiliation{Department of Physics, Duke University, Durham, North Carolina 27708, USA}
\author{P.~U.~Sauer}
\affiliation{Institut f\"ur Theoretische Physik, Leibniz Universit\"at Hannover, D-30167 Hannover, Germany}
\author{R.~Skibi\'nski}
\affiliation{M. Smoluchowski Institute of Physics, Jagiellonian University, PL-30059 Krak\'{o}w, Poland}
\author{A.~P.~Smith}
\affiliation{Triangle Universities Nuclear Laboratory, Durham, North Carolina 27708, USA}
\affiliation{Department of Physics, Duke University, Durham, North Carolina 27708, USA}
\author{M.~B.~Tsang}
\affiliation{NSCL, Michigan State University, East Lansing, Michigan 48824, USA}
\author{H.~R.~Weller}
\affiliation{Triangle Universities Nuclear Laboratory, Durham, North Carolina 27708, USA}
\affiliation{Department of Physics, Duke University, Durham, North Carolina 27708, USA}
\author{H.~Wita{\l}a}
\affiliation{M. Smoluchowski Institute of Physics, Jagiellonian University, PL-30059 Krak\'{o}w, Poland}
\author{Y.~K.~Wu}
\affiliation{Triangle Universities Nuclear Laboratory, Durham, North Carolina 27708, USA}
\affiliation{Department of Physics, Duke University, Durham, North Carolina 27708, USA}
\author{Z.~W.~Zhao}
\affiliation{Triangle Universities Nuclear Laboratory, Durham, North Carolina 27708, USA}
\affiliation{Department of Physics, Duke University, Durham, North Carolina 27708, USA}

\date{\today}

\begin{abstract}
The first measurement of the $\vec{^3He}(\vec{\gamma},p)d$ process was performed at the High Intensity $\gamma$-ray Source (HI$\gamma$S) facility at Triangle Universities Nuclear Laboratory (TUNL) using a circularly polarized,  monoenergetic $\gamma$-ray beam and a longitudinally polarized $^3$He target. The spin-dependent asymmetry and the contribution from the two-body photodisintegration to the $^3$He Gerasimov-Drell-Hearn integrand are extracted and compared with state-of-the-art three-nucleon system calculations at the incident photon energy of 29.0  MeV. The data are in general agreement with the various theoretical predictions based on the Siegert theorem or on explicit inclusion of meson-exchange currents.
\end{abstract}


\maketitle
The study of three-nucleon systems has been of fundamental importance to nuclear physics~\cite{glockle1,Carlson}, and essential to the study of the partonic structure of the nuclei where the $^3$He and $^3$H mirror nuclei are used to extract the ratio of the inelastic structure functions, $\frac{F^n_2}{F^p_2}$~\cite{Marathon}. 
A polarized $^3$He nucleus is often treated approximately as a polarized neutron because its ground state is dominated by the $S-$wave in which the spins of the two protons pair off. Polarized $^3$He targets have been used for decades to extract the electromagnetic form factors~\cite{Gao,Xu,Riordan} and the spin structure functions~\cite{Anthony,Zheng2004} of the neutron, and most recently its three-dimensional structure and dynamics~\cite{Qian2011}. To extract the neutron information from $^3$He, corrections for nuclear effects relying on the state-of-the-art three-body calculations need to be applied. Theoretical calculations using Faddeev~\cite{fad} and Alt-Grassberger-Sandhas equations (AGS)~\cite{Alt} have been carried out for three-body systems using a variety of nucleon-nucleon (NN) potentials~\cite{Stoks,machle,wiringa}, and three-nucleon forces (3NFs) like Urbana IX (UIX)~\cite{pudliner} or CD Bonn + $\Delta$~\cite{Deltuva0}.
It is important to validate these calculations by experiments employing polarized $^3$He nuclei. Data from electrodisintegration of polarized $^3$He~\cite{Xiong} were used to test three-body calculations~\cite{golak}, and more recently data from $\vec{^3He}(\vec{\gamma},n)pp$ channel at incident photon energies of 12.8, 14.7 and 16.5 MeV were reported~\cite{George_PRL,George_PRC,George_PLB,laskaris}. 

Calculations for the two- and three-body photodisintegration of $^3$He with double polarizations have been carried out by two groups. The calculations by Deltuva {\it et al.} are based on the AGS version of Faddeev equations and employ the CD Bonn + $\Delta$ potential~\cite{Deltuva0} taking into account the corresponding single-baryon and meson-exchange electromagnetic currents (MEC). 
The MEC are included in two approaches:
(i) calculating the most important MEC explicitly;
(ii) a more complete approach, including  the dominant part of MEC for electric multipoles implicitly via the Siegert theorem, and the remaining 
part of MEC explicitly, with the Siegert operator including also relativistic single-nucleon charge corrections. In both approaches, the results are obtained using the computational technology of Ref.~\cite{Deltuva2} and the proton-proton Coulomb force is taken into account via the method of screening and renormalization~\cite{Deltuva3}.
Skibi\'nski {\it et al.} solve the Faddeev equations by using the AV18 potential~\cite{wiringa} and the UIX 3NF~\cite{pudliner} with two approaches for MEC:
(i) "pion-in-flight" and "seagull" terms -- the two dominant components of MEC -- are taken into account explicitly~\cite{golak1}; (ii) the dominant MEC contribution to electric multipoles is included implicitly via the Siegert theorem, but only  the nonrelativistic single-nucleon current is considered explicitly.
Their results are obtained using the computational methods described in Ref.~\cite{Skibinski}.
Note that the approaches based on explicit MEC by both groups have quite similar dynamic content,
while in the case of Siegert approaches the included currents are more different.

Another interesting aspect concerning polarized photodisintegration of $^3$He is related to the Gerasimov-Drell-Hearn (GDH) sum rule~\cite{Drell}. The GDH sum rule relates the energy-weighted difference of the spin-dependent total photoabsorption cross sections for target spin and beam helicity parallel ($\sigma^P$ ) and anti-parallel ($\sigma^A$) to the anomalous magnetic moment of the target (nuclei or nucleons) as follows:

\begin{equation}
I^{GDH} = \int_{\nu _{thr}}^{\infty}(\sigma^P- \sigma^A)
{\frac{d\nu}{\nu}} = \frac{4\pi^{2}\alpha}{M^{2}}\kappa^{2} S,
\label{Igdhr}
\end{equation}
where $\nu$ is the photon energy, $\nu_ {thr}$ is the pion production (two-body break-up) threshold on the nucleon (nucleus), $\kappa$ is the anomalous magnetic moment, $M$ is the mass and $S$ is the spin of the nucleon or the nucleus. In $^3$He and below the pion production threshold, only the two-body and three-body photodisintegration channels contribute to the GDH integral with  calculations~\cite{Deltuva2,Skibinski} showing that the three-body channel dominates the integrand. The GDH integrand extracted from measurements of $\vec{^3He}(\vec{\gamma},n)pp$ channel at 12.8 and 14.7 MeV~\cite{George_PRL,George_PRC} is in good agreement with theoretical predictions of~\cite{Deltuva2}, and the result at 16.5 MeV~\cite{George_PLB} is slightly more than one standard deviation higher than the theory.

To fully test the theoretical predictions, not only measurements at higher energies of the three-body break-up channel will be useful. It is also important to test the calculations of two-body breakup channel with double polarizations. A spin-dependent study of $\vec{^3He}(\vec{\gamma},p)d$ reaction together with the $\vec{^3He}(\vec{\gamma},n)pp$ channel will provide stringent tests of the modern three-body calculations, and also serve as an important step towards an experimental test of the GDH sum rule on the $^3$He nucleus by combining inclusive electron scattering measurements above the pion production threshold from other laboratories~\cite{Amarian}.

Experimentally the study of $\vec{^3He}(\vec{\gamma},p)d$ reaction is more challenging than the $\vec{^3He}(\vec{\gamma},n)pp$ channel, especially at low energies due to the necessity of detecting low-energy protons. Such protons are detected in a high background environment from other breakup channels from various nuclear species contained in the $^3$He target wall material. Furthermore, the predicted spin-dependence in the $\vec{^3He}(\vec{\gamma},p)d$ reaction cross section is significantly smaller than that of the three-body channel. As such the experimental study of the $\vec{^3He}(\vec{\gamma},p)d$ channel lags behind the corresponding three-body channel.    

In this Letter, we present the first measurement of the $\vec{^3He}(\vec{\gamma},p)d$ channel using a longitudinally-polarized $^3$He target and the nearly monoenergetic, $\sim$100\% circularly-polarized $\gamma$-ray beam of HI$\gamma$S facility~\cite{higsreview} at $\nu$=29 MeV. The beam intensity on target was 1-3$\times10^{7}\gamma/s$ having an energy spread of 5.0\% (FWHM). A 10.56 cm long C$_6$D$_6$ cell and two BC-501A-based liquid scintillator neutron detectors placed transverse to the beam direction were utilized to measure the photon flux by detecting the neutrons from the deuteron photodisintegration process. The integrated photon flux was extracted based on the well-known cross sections~\cite{bernabei,blackston,birenbaum,graeve,skopik}.

The experimental apparatus used for this measurement comprised two subsystems: the polarized $^3$He target and the detector system. The $^3$He gas target was contained in a one-piece Sol-Gel coated~\cite{Brinker} Pyrex glassware, consisting of a spherical pumping chamber 8.1 cm in diameter and a cylindrical target chamber 39.6 cm long and 2.9 cm in diameter. The two chambers were connected by a transfer tube 0.8 cm in diameter and 9.6 cm long. The target chamber glass thickness was measured using two independent methods, laser interferometry and an ultrasonic gauge, and it was found to vary from $\sim$1.1 mm at the center of the target chamber to $\sim$1.4 mm towards the beam entrance and exit windows. The target was filled with 6.5$\pm$0.1 amg of $^3$He.

The outgoing protons from the $^3$He photodisintegrations were detected by 72 fully depleted silicon surface barrier detectors placed at the proton scattering angles of $45\,^{\circ}$, $70\,^{\circ}$, $95\,^{\circ}$ and $120\,^{\circ}$ degrees (18 detectors at each angle). Six aluminum hemispheres were used to place the detectors $\sim$10 cm away from the center of the $^3$He target chamber (three hemispheres on each side of the cell facing each other). Each hemisphere housed twelve detectors in and out of plane at various scattering angles around the $^3$He target chamber. Collimators with rectangular apertures of 2 cm $\times$ 0.4 cm and a length of 3 cm were placed in front of the detectors. The detector thicknesses ranged from 300 to 500 $\mu$m, and their efficiency for detecting charged particles was 100\%.

The spin exchange optical pumping technique~\cite{Happer} was used to polarize $^3$He target. A small quantity of Rb and K mixture was placed inside the pumping chamber which was heated to 196 C$\,^{\circ}$. A circularly-polarized 794.8 nm laser light incident on the pumping chamber polarized Rb atoms which in turn transferred their polarization to $^3$He nuclei through spin-exchange collisions between Rb-K, Rb-$^3$He and K-$^3$He. A small quantity of N$_{2}$ (0.1 amg) was added into the cell as a buffer gas to improve the optical pumping efficiency. A pair of Helmholtz coils $\sim$170 cm in diameter providing a 20 G magnetic field was used to define the direction of the $^3$He nuclear polarization. The spin of the target was flipped every 15 min. The nuclear magnetic resonance-adiabatic fast passage ~\cite{Lorenzon} calibrated by the electron paramagnetic resonance technique~\cite{epr4} was employed to measure the absolute target polarization. While a polarization over 40\% from target named "SPOT" was achieved in the three-body photodisintegration experiment~\cite{George_PRL,George_PRC} and a 35\% polarization in a follow-up experiment~\cite{George_PLB}, for this two-body breakup measurement the polarization of the target was measured to be 33\% for the initial run period ($\sim$40\% of the beam time) and 22\% for the final runs ($\sim$60\% of the beam time) due to some hardware failure during the experiment. More details about this target can be found in~\cite{laskaris,Kramer,Ye}. 

\begin{figure}[!ht]
  \centering
    \includegraphics[width=0.50\textwidth]{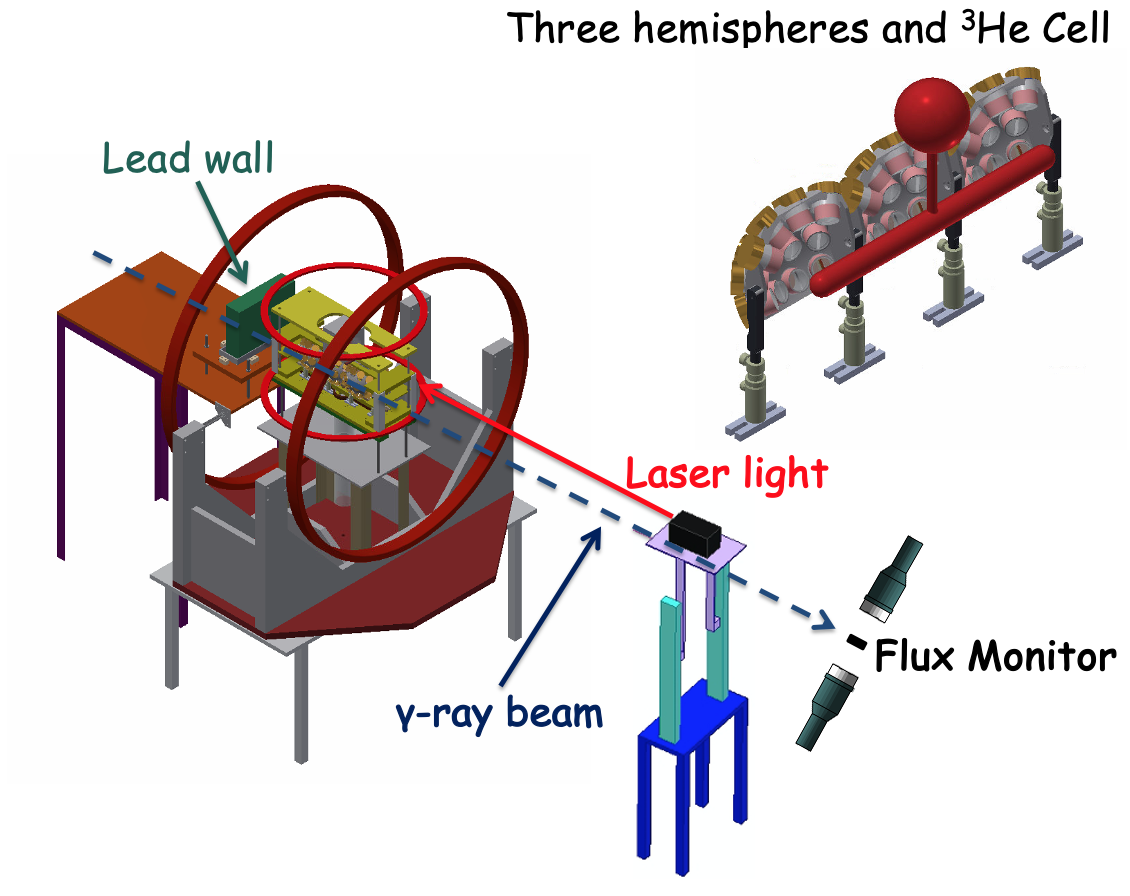}
    \caption{(Color online) A view of the experimental apparatus (not to scale). The movable target system that was used to cycle between the $^3$He target and the N$_2$ reference cell is surrounded by 72 silicon surface barrier detectors. Half of the detector system (three hemispheres supporting 36 detectors) and the $^3$He cell can be seen at the top right. The lead wall placed in front of the targets and the detector system can be seen at the left. The movable support of the laser system used to polarize $^3$He can be seen next to the C$_6$D$_6$ flux monitor at the right.}
    \label{fig:experimental_apparatus}
\end{figure}

A N$_2$-only reference cell with the same dimensions as those of the $^3$He target chamber was filled with the same amount of N$_2$ gas and placed right-below the $^3$He target to measure backgrounds. A lead wall with a 16 mm aperture allowing for the $\gamma$-beam to pass was placed in front of the targets and the detector system to attenuate the beam halo and reduce the electron background. Figure~\ref{fig:experimental_apparatus} shows a schematic view of the experimental apparatus including the polarized $^3$He target subsystem, the 72 detector subsystem and the C$_6$D$_6$ flux monitor.

Two quantities were recorded for each event, the incident charged particle energy and the relative time of flight (TOF) between the silicon detectors and the RF signal of the beam. A two-dimensional cut was applied to the energy plotted against the relative TOF and used to select the protons from the $^3$He cell. The same cuts were applied to the data taken with the N$_2$ reference cell to subtract the proton background from other processes. The protons from the competing $\vec{{^3}He}(\vec{\gamma},p)pn$ reaction could not be subtracted using the N$_2$ reference cell. However, a GEANT4~\cite{geant} simulation using the measured glass thicknesses of the target cell has shown that no protons from the three-body photodisintegration of $^3$He can make into the detectors. After selecting the protons, the spin-dependent asymmetry for each detector can be formed as 

\begin{equation}
A = \frac{1}{P_bP_t}\frac{Y^P-Y^A}{Y^P+Y^A}
\label{Ia}
\end{equation}
where P$_b$ and P$_t$ are the beam and target polarization, respectively, and Y$^{P/A}$ are the integrated normalized yields (proton counts/integrated photon flux) with $Y^{P/A}=Y^{P/A,^3He}-Y^{N_2}$ being the measured yield from the $^3$He cell after the subtraction of the N$_2$ reference cell background yield for both parallel and anti-parallel states. Although the uneven glass thickness of the $^3$He target chamber affected the proton yields for each detector, it did not affect the asymmetry as was shown by a GEANT4~\cite{geant} simulation of the experiment. This allowed the calculation of the asymmetry for each angle as the weighted average of the asymmetries of all 18 detectors at this angle. While many systematic uncertainties cancel by forming the asymmetry, two remaining contributions to the systematic uncertainty are the target polarization of 4.2\% and the beam polarization of 1.0\% resulting to an overall relative systematic uncertainty of 4.3\%.

\begin{figure}[!ht]
  \centering
    \includegraphics[width=0.50\textwidth]{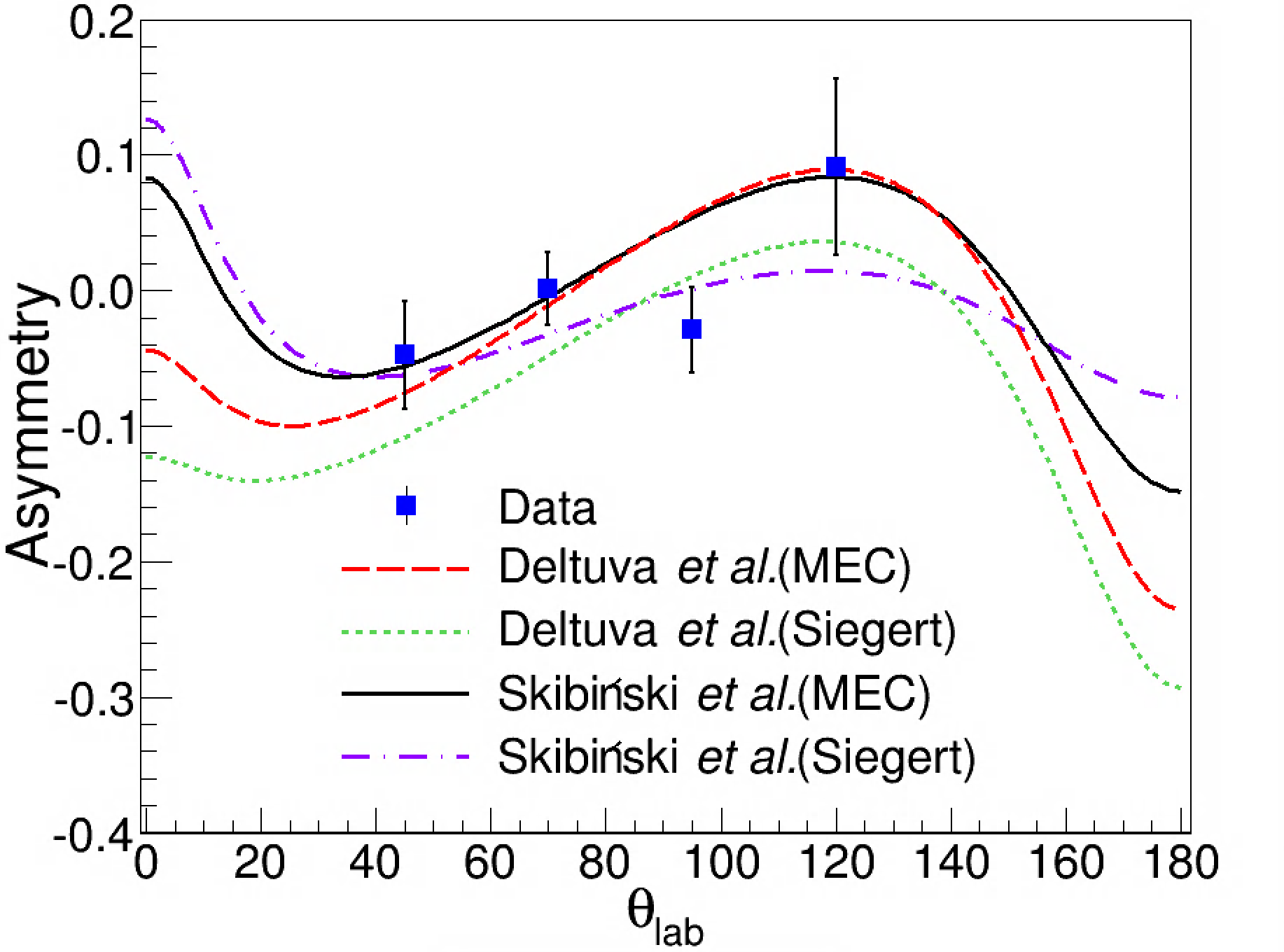}
    \caption{(Color online) Measured spin-dependent asymmetry including statistical and systematic uncertainties compared with the calculations of Deltuva {\it et al.}~\cite{Deltuva2} and Skibi\'nski {\it et al.}~\cite{Skibinski} at $\nu$=29.0 MeV.}
    \label{fig:asymmetry}
\end{figure}

The measured spin-dependent asymmetries as a function of the proton scattering angle, $\theta_{lab}$ at $\nu$=29 MeV are shown in fig.~\ref{fig:asymmetry} including statistical and systematic uncertainties added in quadrature. The data are compared with the two sets of theoretical calculations provided by Deltuva {\it et al.}~\cite{Deltuva2} and Skibi\'nski {\it et al.}~\cite{Skibinski}. Although the calculations based on the Siegert theorem with relativistic charge corrections are considered to be more complete, the overall shape of the experimental results seem to be described better by the calculations taking into account the MEC explicitly. However, one can not reach a definitive conclusion as to which theoretical calculation is favored by the asymmetry data given the overall uncertainties. 

By combining the measured asymmetry, the known angular distribution of the unpolarized differential cross sections~\cite{anghinolfi} and the total cross sections~\cite{kundu,ticcioni} at 29 MeV, one can extract the spin-dependent differential cross sections. Second order Legendre polynomials are used to fit the spin-dependent differential cross sections and the fitting curves are integrated over the angle to extract the spin-dependent total cross sections and the GDH integrand. 

\begin{table}[!h]
\begin{center}
\caption{The extracted spin-dependent total cross sections, $\sigma^{P}$ and $\sigma^{A}$, and the contributions from the two-body photodisintegration to $^3$He GDH integrand, $(\sigma^{P}-\sigma^{A})/\nu$ compared with theoretical predictions.}
\begin{tabular*}{0.50\textwidth}{@{\extracolsep{\fill} }l c c c }
\hline
\hline
 & $\sigma^{P}$($\mu$b) & $\sigma^{A}$($\mu$b) & $(\sigma^{P}-\sigma^{A})/\nu$ (fm$^{3}$) \\ 
\hline
This work & 277$\pm$32 & 276$\pm$30 & 0.00068$\pm$0.02769  \\
Deltuva {\it et al.} (MEC) & 305 & 306 & -0.00068 \\
Deltuva {\it et al.} (Siegert) & 309 & 336 & -0.0184 \\
Skibi\'nski {\it et al.} (MEC) & 303  & 299 & 0.00272 \\
Skibi\'nski {\it et al.} (Siegert) & 295  & 310 & -0.0102 \\
\hline
\hline
\end{tabular*}
\label{table:gdh}
\end{center}
\end{table}

Table~\ref{table:gdh} summarizes the extracted spin-dependent total cross sections and the contribution from the two-body photodisintegration to the $^3$He GDH integrand in comparison to the two sets of calculations from Deltuva {\it et al.}~\cite{Deltuva2} and Skibi\'nski {\it et al.}~\cite{Skibinski}. The reported uncertainties include the statistical and systematic uncertainties of the current asymmetry measurement and the uncertainties associated with the known angular distribution~\cite{anghinolfi} and the total cross sections~\cite{kundu,ticcioni}. The extracted spin-dependent total cross sections are slightly smaller in magnitude but within $\sim$1-$\sigma$ from the calculations. As expected based on the asymmetry results, the extracted GDH integrand seems to favor the explicit MEC-based calculations.

\begin{figure}[!ht]
  \centering
    \includegraphics[width=0.50\textwidth]{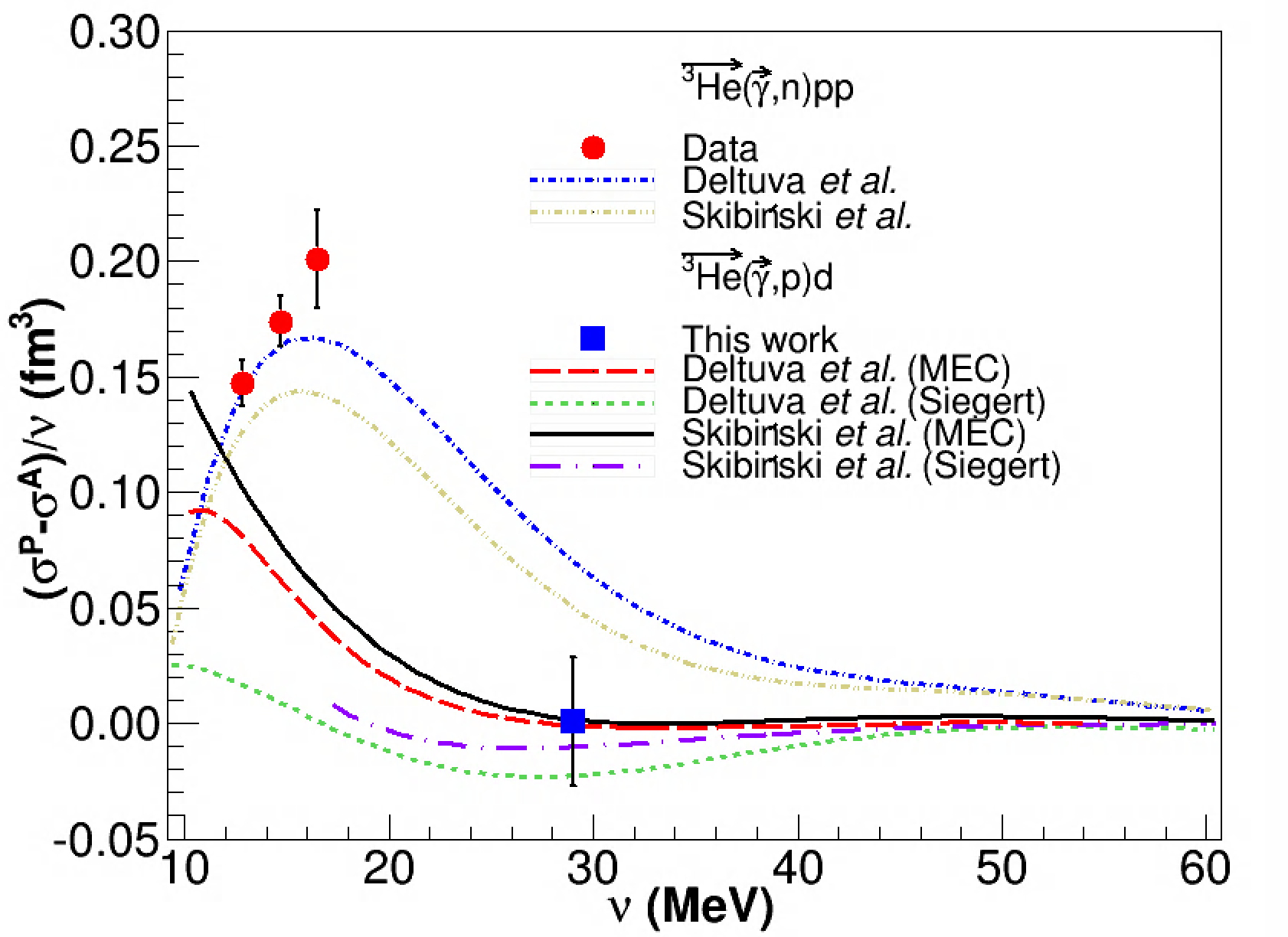}
    \caption{(Color online) The extracted GDH integrand of $\vec{{^3}He}(\vec{\gamma},p)d$ (blue square) plotted together with the results from $\vec{{^3}He}(\vec{\gamma},n)pp$ (red circles)~\cite{George_PRL,George_PLB,George_PRC,laskaris} including statistical and systematic uncertainties compared with the calculations of Deltuva {\it et al.}~\cite{Deltuva2} and Skibi\'nski {\it et al.}~\cite{Skibinski}.}
    \label{fig:GDHIntegrand}
\end{figure}

Fig.~\ref{fig:GDHIntegrand} shows the contributions from two-body photodisintegration to the $^3$He GDH integrand together with the two sets of predictions from Deltuva {\it et al.}~\cite{Deltuva2} and Skibi\'nski {\it et al.}~\cite{Skibinski} as a function of the incident photon energy. In the same figure the past measurements of the contributions from the three-body photodisintegration to the $^3$He GDH integrand together with the predictions from Refs.~\cite{Deltuva2} and~\cite{Skibinski} are shown for comparison. Noteworthy, the best description of two- and three-body data is given by different calculations, based either on explicit MEC or Siegert with relativistic charge corrections, respectively. 

In this Letter, we report the first measurement of the double polarized $\vec{{^3}He}(\vec{\gamma},p)d$ reaction. It is remarkable to note that the new data and the previous data on the three-body channel support the theoretical predictions of the dominance of the $\vec{^3He}(\vec{\gamma},n)pp$ channel over the $\vec{^3He}(\vec{\gamma},p)d$ channel in the contribution to the $^3$He GDH integrand in this low-energy region. Providing additional data for the observables sensitive to the details of exchange currents is important in view of future analysis with chiral currents. These additional data, when combined with data from three-body photodisintegration and data above pion production threshold from other laboratories, will directly test the theoretical calculations and the $^3$He GDH sum rule predictions.

The authors would like to dedicate this letter to the memory of Adamos Kafkarkou, a brilliant young physicist who left this world early. We would also like to thank M. Souza and G. Cates for their assistance in constructing the Sol-Gel coated cell and the TUNL personnel, in particular, the HI$\gamma$S operation team for the technical support of this experiment. This work is supported by the U.S. Department of Energy under contract numbers DE-FG02-03ER41231, DE-FG02-97ER41033, DE-FG02-97ER41041, DE-SC0005367, the US National  Science  Foundation under contract number  PHY-1565546, and the Polish National Science Center under Grant No. 2016/22/M/ST2/00173. A.D. acknowledges support by the Alexander von Humboldt Foundation under Grant No. LTU-1185721-HFST-E. The numerical calculations of Krak\'ow theoretical group have been performed on the supercomputer clusters of the JSC, J\"ulich, Germany.

\end{document}